# USING PARALLEL PROCESSING FOR FILE CARVING

Nebojša Škrbina
European University
Skopje, Republic of Macedonia

Toni Stojanovski
European University
Skopje, Republic of Macedonia

### ABSTRACT

File carving is one of the most important procedures in Digital Forensic Investigation (DFI). But it is also requires the most computational resources. Parallel processing on Graphics Processing Units have proven to be many times faster than when executed on standard CPU. This paper is inspecting the algorithms and methods to use parallel processing for development of file carving tools that will do their job much faster than the conventional DFI tools.

Keywords: Parallel Processing, Digital Forensic Investigation, File Carving, GPGPU, String Search Algorithms

## I. INTRODUCTION

Parallel processing is type of programming which is designed to be executed on processors with parallel architecture. Such parallel architecture is implemented in current generation of Graphics Processing Units (GPUs). Traditionally, GPUs have been designed to perform very specific type of calculations on textures and primitive geometric objects, and it was very difficult to perform non-graphical operations on them. One of the first successful implementation of parallel processing on GPU using non-graphical data was registered in 2003 [1]. Latest generations of graphics cards incorporate new generations of GPUs, which are designed to perform both graphical and non-graphical operations. One of the pioneers in this area is NVIDIA [2] with CUDA (Compute Unified Device Architecture). Others also have their own programmable interfaces for their technologies: AMD ATI Stream, Open CL etc.

Today GPU-based parallel programming is omnipresent: biology, genetics, physics, economics [3], [4], [5] etc. GPUs are also extensively used in digital forensics [5], [6], [7]. Digital forensic is relatively new discipline within computer science and deals with discovery of evidence on digital devices. Digital Forensic Investigation (DFI) has to answer several questions:

- What events have happened?
- When did they happen?
- Who was responsible?
- What was the mechanism by which the events occurred?

According to Carrier [5], "Digital evidence is data that supports or refutes a hypothesis that was formulated during the investigation. This is a general notion of evidence and may not be court admissible because it was not properly or legally acquired". The whole process of DFI must be conducted effectively and independently, because in some cases, someone's life or liberty can be jeopardized.

File carving is technique used in DFI for recovering files when there is no file system existing on the disc. It uses binary and string search algorithms to search for data patterns that can identify files. File carving is a computationally intensive operation. More precisely, processing is simple but it has to be done over huge amounts of data. Hence, file carving is one of the potential areas of application for Parallel Processing.

Our research addresses the following questions:

- How to implement CUDA technology in DFI?
- What is the difference between string matching and pattern matching algorithms and which one is better for using in parallel processing?
- What are the differences in implementation of file carving method for different file systems (FAT, NTFS, EXT)?
- Which algorithm is most appropriate for parallel implementation in file carving?
- What is the speed-up for the selected algorithm for file carving when executed on GPU compared to CPU?

As a contribution, this paper answers the above questions, and gives a comparative analysis of the applicability of binary search methods in parallel processing.

Here is the overview of the paper. Section II gives an overview of the areas of GP-GPU, Digital Forensics, File Carving and related work. Section III contains the main results of the paper. Section IV concludes the paper, and gives directions for future research.

## II. OVERVIEW

### A. GPGPU

In the last decade, computer games have evolved into complete multimedia experience with very complex and realistic graphical capabilities. In order to satisfy market demands for high definition 3D graphics, GPUs have evolved into a highly parallel, multicore processor with big computational power and high memory bandwidth.

For example, the latest product from NVIDIA called GeForce GTX 680 based on new 28nm "Kepler" core architecture has eight Streaming Multiprocessor (SMX) units consisting of 1536 CUDA cores [8]. Theoretical computing performance is 3090 GFLOP/s with suggested retail price of $499 (March 2012). The most powerful "classical" CPU from Intel is model i7-3960 on 3.3 GHz containing 6 cores with retail price of 1100$ (March 2012) [9]. Theoretical computing performance for this Extreme Edition Series processor is 158 GFLOP/s. Historical differences in performance between CPU and GPU are presented in Figure 1 [10].

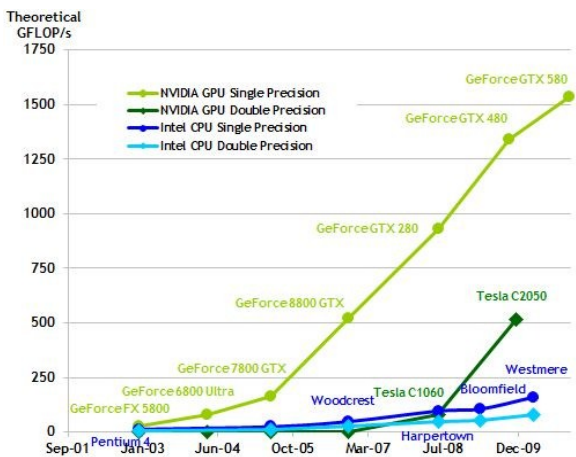

Figure 1. GFLOP/s for CPU and GPU.

This big difference in GFLOP/s between CPU and GPU is because GPU is designed for intensive, highly parallel computation, exactly for graphic rendering purposes. In GPUs architecture much more transistors are devoted to data processing and less to data cashing, as illustrated by Figure 2 [10].

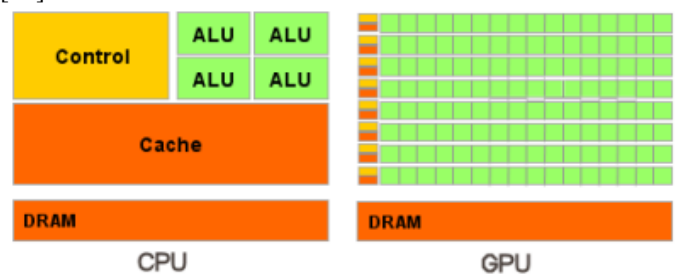

Figure 2. Difference in CPUs and GPUs architecture.

It would be waste of power if these GPUs capabilities were not used for non-graphical computation. In the last decade a lot of research has been devoted to general purpose computing on GPUs, collectively known as GPGPU [4]. Many of those researches were conducted on NVIDIA, one of the world's leaders in research and development of Graphics Processing Units. Their technology is called CUDA. It is software and hardware architecture for managing computations on GPU without mapping them on graphics API. We are using CUDA technology for our research.

*B. Digital Forensic Investigation (DFI)*

DFI has many characteristics, but it must comply with the following four requests. Effective DFI must be: reliable, comprehensive, efficient and coherent. Investigation is *reliable* when evidence is accurate and free from tampering. *Comprehensive* investigation should analyze as many (if not all) potentially interesting targets as possible. *Efficient* investigation is when it uses maximum of all available resources like computer power, time, available storage and man power. Combining parallel processing in tools for digital forensic will improve the efficiency of investigation. The purpose of our research is to improve the before mentioned process. To conduct coherent investigation is probably the most important request. Investigation must provide evidence from the analysis that can be used to compile an integrated view of the events under question.

DFI process consists of several phases. The first phase is collecting the evidence from digital devices. That device can be any device which is used for storing data: HDD, USB flash memory, USB HDD, CD, DVD, PDA, memory in mobile phones and/or any kind of memory card (external or internal), storage spaces in clouds, etc. The second phase is examination of data, which is called "digital evidence". This is the most time consuming phase, and our research aims to help make this phase faster and more reliable. The third phase is analysis of gathered digital evidence, and the last phase is reporting about the whole process. It is important to emphasize that phases are not very strictly divided in the time schedule, and the researcher is allowed to repeat or to search for more data in any phase if necessary. The whole process is shown in Figure 3 [7].

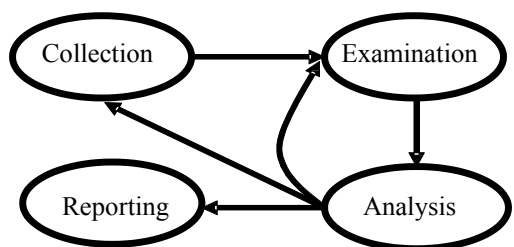

Figure 3. DFI Process.

DFI must be conducted within the constraints of processing power, storage capacity and available time. Managing all these limitations can be very difficult, especially when DFI is conducted in the field. If one or two of these limitations are not adequate the third one can compensate for them, but within some limitations. For example, if available computer power is limited, it can be compensated with extended time scheduled for that part of investigation. Or, large amount of data can be processed faster with stronger computer power. In order to understand how big DFI cases are today, we present following examples. In 2003, USA Department of Defence conducted investigation about leakage of plans for operations in Iraq and the amount of data was over 60TB [11]. In the investigation regarding Enron fiasco, 31TB of data were analyzed. In Table 1, taken from the FBI [12], we can see increasing amount of data in average case size over the last eight years.

|  | Examinations | Data Processed | Average Case Size |
| --- | --- | --- | --- |
| FY 2003 | 987 | 82 TB | 83 GB |
| FY 2004 | 1304 | 229 TB | 175 GB |
| FY 2005 | 2977 | 457 TB | 153 GB |
| FY 2006 | 3633 | 916 TB | 252 GB |
| FY 2007 | 4634 | 1288 TB | 277 GB |
| FY 2008 | 4524 | 1756 TB | 388 GB |
| FY 2009 | 6016 | 2334 TB | 387 GB |
| FY 2010 | 6564 | 3086 TB | 470 GB |

Table 1. Case trends

*C. File carving*

Having reliable data recovery procedure is very important for every firm, from user to server level. Traditional data recovery is based on file system structure to recover data. Most file systems do not delete the file; they delete information about that file in file structure. Therefore, it is possible to extract information from file structure and to recover deleted file. Different operating systems use different file systems:

- Windows (FAT 12/16/32, NTFS)
- Linux (Ext2/Ext3/Ext4, Reiser)
- Mac (HFS, HFS+/HFSX)

Carrier [13] describes that "file system consists of structural and user data that are organized such that the computer knows where to find them." But, when file system or structure is damaged or completely missing, it is not possible to use any traditional data recovery methods. One of the techniques for data recovery is called *file carving*. It is method for recovering files which is based on the content of media storage (most usually it is HDD) [14]. It uses knowledge about the structure of the file. In order to understand file carving process, an understanding of file systems and fragmentation is required. The smallest addressable space (can be written to or read) on disc is cluster. Cluster size is from 512B to 32000B, depending on the file system type and the available space on the device storage. Therefore, it is important to understand that files are stored in clusters.

There are many techniques used in file carving [15], but the most common are:

- Header-footer or header – "max file size" carving
- File structure based carving
- Content based carving

Our research is based on header-footer file carving illustrated in Figure 4. Every file has beginning (header) and end of file (footer) which is determinate by file type. Searching for these well known signatures can determine where the file starts and where it ends, but only if it is not fragmented. If footer cannot be found, then header – "max file size" technique is used.

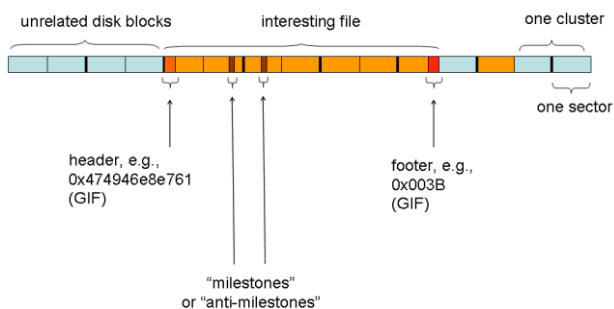

Figure 4. Header – footer technique of file carving

File structure based carving uses the internal layout of a file like header, footer, size information and some identifier strings. Programs that use this technique are Foremost [16] and PhotoRec [17]. Content based carving uses information about content structure like: character count, text / language recognition, statistical attributes, information entropy etc.

*D. Related work*

There are several works that are similar to ours. The work most related to ours is the work by Lodovico Marziale et al. [6]. It presents the results of several experiments that evaluate the effectiveness of offloading processes common to digital forensic tool to a GPU. In these experiments, "massive" numbers of threads (65536) are used to parallelize the computation. Their results indicate that there are significant increases of performance of digital forensic tools when they are designed and executed on GPUs. From the same author is [7], which is his PhD work, continuing his previous work in more details.

In [18] Giorgos Vasiliadis et al. have implemented IDS (Intrusion Detection System) called Gnort to run on GPU. They have used pattern matching method in combination with Aho – Corasick algorithm to increase performance of IDS system by factor of two. Same author in [19] presents design, implementation and evaluation of regular expression matching engine which is executed on GPU. They have implemented this method in popular Snort IDS [20] and gain increase in package processing throughput by 60%.

Bai Hong-tao et al. [21] present K – means algorithm based on SIMD (Single Instruction Multiple Data) architecture optimized for GPUs. Their numerical experiments demonstrated that the speed of GPU – based K – means could reach as high as 40 times of the CPU – based K – means.

In [22], Peter J. Lu et al. are presenting implementation of image correlation on GPU using NVIDIA's CUDA platform. They are using their own code for analyzing images of liquid gas phase separation, photographed in zero gravity aboard the International Space Station. That GPU code is 4000 times faster than MATHLAB code performing the same calculation on CPU.

Chris Messom and Andre Barczak in [23] present design and evaluation of the stream processing implementation of the Integral Image algorithm. That algorithm is the key component of many image processing algorithms, in particular the Haar – like feature based systems. This research results in significant performance improvement when calculations are executed on GPU using CUDA technology.

III. MAIN RESULTS

*A. String search algorithms*

Arguably the most important decision in our research is choosing the right algorithm for parallel implementation on GPU using CUDA. There are many candidates for this job. Some of their parallel implementations are presented in [6], [18], [19], [20] and [24]. String search algorithms can be divided into *single string* and *multi string* search algorithms.

*Single string* searching algorithm is designed to search one string (pattern, text, etc.) at a time. It means that if we have to search for *m* strings, the whole algorithm must be repeated *m* times over the whole data base (in our case it is the entire HDD). One of the oldest, most popular and slowest algorithms is brute force search. Some of the most widely used single string searching algorithms are Knut – Morris – Pratt (KMP) [25] and Boyer – Moore (BM) [26]. KMP algorithm builds a preprocessing table for every string separately. During the search (comparison) phase it uses the preprocessed table to skip characters if a mismatch happens. BM algorithm searches for a string from right to left and if a mismatch happens, it shifts to the right for the length of that string. Single string searching algorithms are very good candidates for parallel processing because they have relatively simple logic and they need large computational power in order to be usable.

*Multi string* search algorithms search for set of strings in database simultaneously. In the preprocessing phase they build a logic to be used in the search phase. That logic can be represented as a decision tree, or a table, or a combination of both. By using this preprocessed logic, every character in the data base will be searched only once. The most popular multi string search algorithms are: Aho – Corasick (AC) [27], Commentz – Walter [28] and Wu – Manber [29]. AC algorithm constructs a finite state machine that looks like a tree with links between internal nodes. Commentz – Walter algorithm is a combination of AC and BM algorithms, while Wu – Manber is improved BM algorithm. Multi string search algorithms are faster than algorithms that search for each pattern individually, when they are executed sequentially on CPU [18].

*B. File signatures*

Every file type has its distinguished structure which is unique for that file type. Beginning of file (header) and end of file (footer) are part of that structure. It is important to emphasize that file structure is not related to file system. For example, PDF file will have the same file structure irrespective of the file system where it is stored. This characteristic is used in header – footer file carving techniques. Table 2 gives headers and footers of some well known files.

| File type | Header | Footer |
|---|---|---|
| JPEG | \xFF\xD8 | \xFF\xD9 |
| GIF | x47\x49\x61 | \x00\x3B |
| ZIP | PK\x03\x04 | \x3C\xAC |
| PDF | %PDF | %EOF |
| PST | !BDN | - |

Table 2. Headers and Footers of files

In the terms of file system, headers and footers are also known as magic numbers [30], because they are constants used to identify a file format. But some files (like MS Office files) do not have footers and in that case it is necessary to conduct file carving with header – "max file size" carving. In these cases deeper knowledge of internal file's structure is needed. For example, in the header of the first sector of office file, there must be hex value FE and FF in the 29$^{th}$ and 30$^{th}$ character respectably. Incorporating deeper knowledge of internal file's structure will decrease number of false positive results in file carving.

*C. Applicability of search algorithms for file carving*

The string matching problem consists of finding one, many or all occurrences of predefined pattern (string) in a text. Searched text can be of any size and type. String searching algorithms have been used in many areas of computer science: Image recognition [22], [23]; Intrusion Detection Systems [18], [19], [24]; Digital Forensic Investigations [7], [6]; and in other fields. Searched strings in file carving process (headers and footers) are relatively small, as shown in Figure 5. Therefore, we believe that string searching algorithms are applicable for this job.

Based on our research and on [4], [7], [6], [18], [19], [25], [26], [27], [28] and [29], we believe that best candidates for file carving by parallel processing are Boyer – Moore and Aho – Corasick algorithms.

Boyer – Moore algorithm is single string searching algorithm that was presented in 1977 by Robert S. Boyer and J. Strother Moore [26]. It is based on the "sublinear" principle, which means that it is not necessary to check every symbol in the searched text. If the pattern is longer and the alphabet is bigger, the algorithm will work faster, because it will be able to skip more symbols in the text. The main feature of this algorithm is its scan logic. It compares searched string with text starting from right side and if match is found, then it moves to the left, until the end of searched string. If mismatch occurs, it shifts to the right for predefined number of characters, according to logic defined by searched pattern. This search logic is presented in Figure 5.

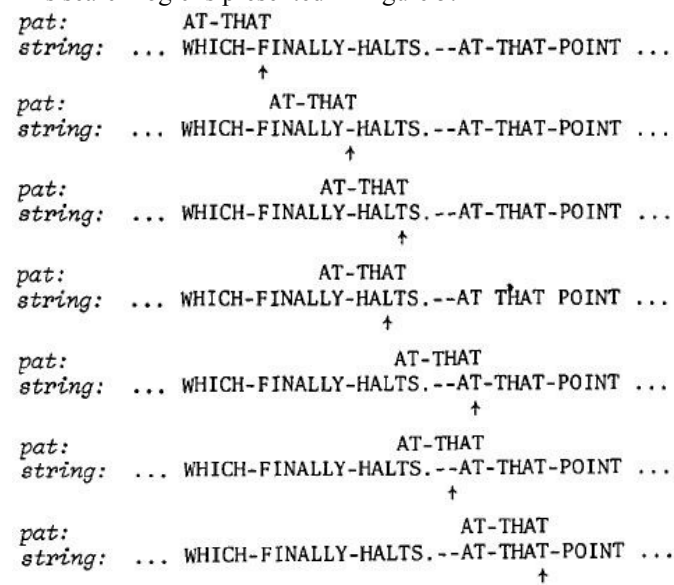

Figure 5. Search logic of Boyer – Moore algorithm

Aho – Corasick algorithm is multi string search algorithm designed by Alfred V. Aho and Margaret J. Corasick in 1975. Their approach combines ideas from KMP algorithm with those of finite state machines [27]. It consists of two parts. In the first part (called preprocessing phase) a finite state pattern matching machine is constructed from the set of keywords (strings). In the second part, that set of keywords is applied as input to the pattern matching machine. Whenever a match is found, machine will signal in a predetermined way. Three functions called *goto*, *failure* and *output* determine the behavior of the pattern matching machine. All searched strings are contained in the *goto* function, and the algorithm runs only ones through the entire database, searching for all strings at the same time. Figure 6 shows *goto* function for following set of keywords: he, she, his and hers. The path 0 – 1 – 2 spells the keyword "he", so "he" will be associated with state 2. Second keyword "she" is represented by path 0 – 3 – 4 – 5, and output "she" is associated with state 5. Keyword "his" is represented by 0 – 1 – 6 – 7 and associated with state 7. And the last one "hers" is 0 – 1 – 2 – 8 – 9 and associated with state 9. For every keyword there is corresponding path and associated state represented in a form of rooted directed tree.

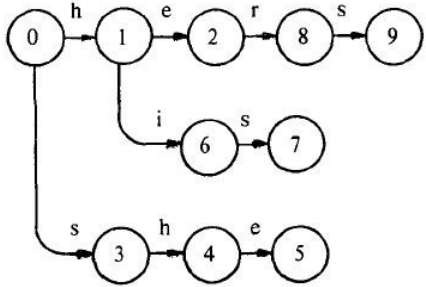

Figure 6. Goto function of Aho – Corasick algorithm

In the parallel processing, the host computer (CPU) is transferring a unit of work to the GPU which is called kernel. That work is executed on GPU in the form of many different threads organized in thread blocks. Every block is executed by one microprocessor in a SIMD principle. The main problem in parallel programming is avoiding *if – then – else* branching, as much as possible, because it leads to thread divergence. Too many divergences will transform parallel processing into sequential processing usually executed on regular CPU.

Our main advantage regarding this problem is the fact that we know in advance what the searched keywords will be. They are already defined headers and footers of well known file types, as shown in Table 2.

IV. CONCLUSION

According to Nicole Beebe [31], more intelligent analytical approach is needed in the future development of DFI tools. In every DFI tool, the percentage of false positives is very high and supervision of skilled investigator is required. But it is hard to find such professionals in larger numbers, and they are often highly paid. The cost of human intervention is significant in every investigation. Reducing human intervention by increasing computational time should be considered as an advantage. We believe that our research will help in reducing that increased computational time by developing more efficient DFI tools that can be used on GPU.